\documentclass[aps,epsfig,showpacs,twocolumn]{revtex4}
\usepackage{amsmath}
\usepackage{amssymb}
\usepackage{graphicx}
\usepackage{subfigure}

\begin{document}

\title{Effect of Quantum Point Contact Measurement on Electron Spin State in
Quantum Dot }
\author{Fei-Yun Zhu$^{(1)}$}
\author{Zhi-Cheng Zhu$^{(1)}$}
\author{Hua Tu$^{(2)}$}
\author{Tao Tu$^{(1)}$}
\email{tutao@ustc.edu.cn}
\author{Guang-Can Guo$^{(1)}$}
\author{Guo-Ping Guo$^{(1)}$}
\email{gpguo@ustc.edu.cn}
\affiliation{$^{(1)}$ Key Laboratory of Quantum Information, University of Science and
Technology of China, CAS, Hefei, 230026, P. R. China\\
$^{(2)}$ Taiping People's Hospital of Dongguan City, Dongguan, 523905, P. R.
China }
\date{\today}

\begin{abstract}
We study the time evolution of two electron spin states in a double
quantum-dot system, which includes a nearby quantum point contact (QPC) as a
measurement device. We obtain that the QPC measurement induced decoherence
is in time scales of microsecond. We also find that the enhanced QPC
measurement will trap the system in its initial spin states, which is
consistent with quantum Zeno effect.
\end{abstract}

\pacs{03.67.Lx, 03.65.Yz, 03.65.Ta} \maketitle

\baselineskip16pt

\textit{Introduction.} Recently, electron spin in semiconductor quantum dot
becomes one of the most promising candidates for quantum computing \cite%
{Loss1998,Hanson2007}. The single electron spin states \cite{Loss1998} or
singlet and triplet states of two electrons in double quantum-dot \cite%
{Taylor2005} have been proposed as qubits in quantum computing. However
these spin qubits are not completely decoupled from the environment, and
there are various sources of decoherence that are intrinsic to the quantum
dot system. For example, the electrons spins couple to phonon in the
surrounding lattice or other fluctuations via the spin-orbit interaction
\cite{Nazarov2000,Nazarov2001,Loss2004} and the electron spins couple to the
surrounding nuclei via the hyperfine interaction \cite{Loss2002,Taylor2007}.
The spin-orbit related relaxation time $T_{1}$ $\approx 1$ $\mathrm{ms}$ has
been demonstrated experimentally \cite{Elzerman2004}. Recently through
spin-echo \cite{Petta2005} or dynamical nuclear polarization technology \cite%
{Reilly2008} to suppress the influence from nuclear spins, the dephasing
time $T_{2}$ $\approx 1$ $\mathrm{\mu s}$ is also achieved in double
quantum-dot system.

On the other hand, as a read-out device, quantum point contact (QPC) is used
as a detector \cite{Hanson2007} to read out spin information in quantum dot
system, meanwhile it has back action to the system \cite%
{Gurvitz1997,Loss2004-2}. It is naturally desired to study how QPC affects
the spin states of the quantum dot via Coulomb interaction during the
measurement process. It is anticipated that the decoherence in the
measurement process would be especially important in case the influence of
nuclear spin is suppressed \cite{Petta2005,Reilly2008} or there is completed
eliminated hyperfine interaction in the SiGe or graphene quantum dots \cite%
{SiGe,Graphene}.

In this paper, firstly, we introduce a model for two electron spins in a
double quantum-dot system with a nearby QPC, give the master equation and
calculate the spin states evolution due to the QPC measurement. We make some
numerical predictions of QPC induced decoherence time for real experimental
parameters. Further, we find that the enhanced QPC measurement will slow
down the transition rate between two spin states, which is consistent with
quantum Zeno effect.

\begin{figure}[ht]
\subfigure [] {\label{1a}\includegraphics[width=0.7\columnwidth]{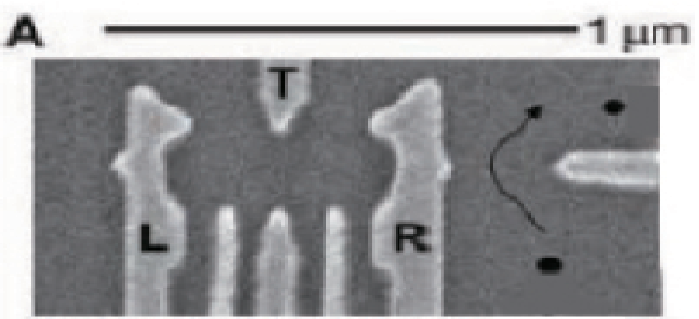}} %
\subfigure[]{\label{1b}\includegraphics[width=0.7\columnwidth]{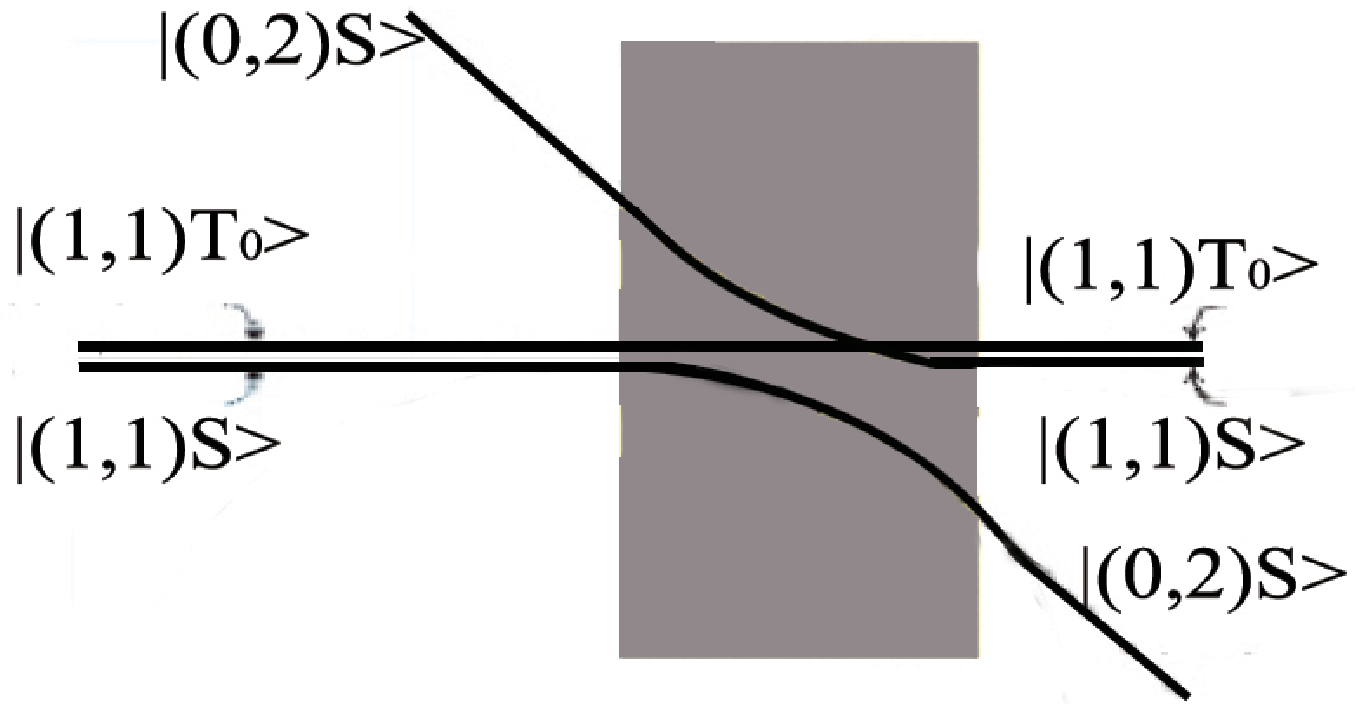}} %
\subfigure[]{\label{1c}\includegraphics[width=0.7\columnwidth]{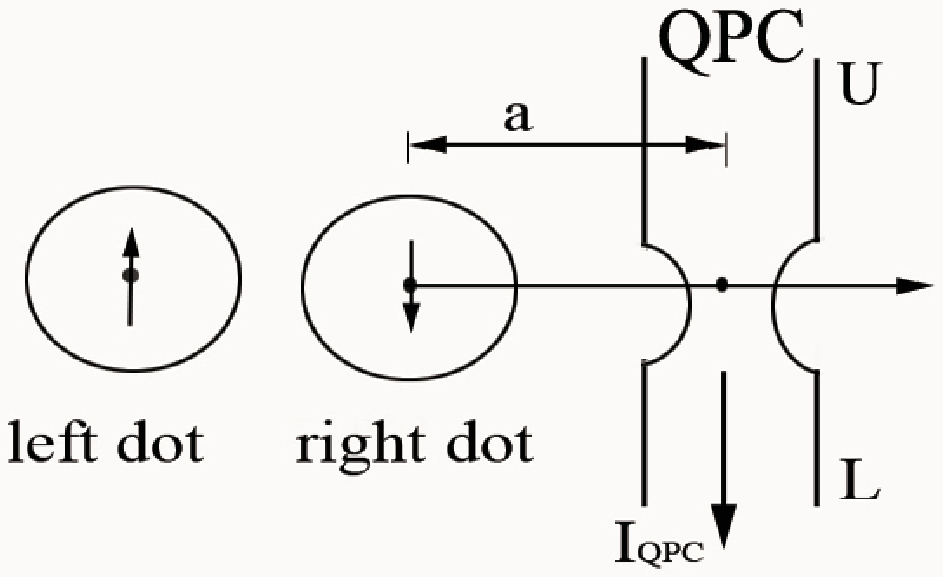}}
\caption{(a) Micrograph of a sample of double quantum dots with a QPC in
close proximity to the right dot. This figure is reproduced from Fig. 1 in
\protect\cite{Petta2005}. (b) Lowest-energy states of the double quantum-dot
system. (c) Demonstration of QPC near the double quantum-dot system. $a$
gives the QPC position.}
\label{fig1}
\end{figure}

\textit{Model and master equation.} Our model is motivated by recent
experiment for manipulation and measurement of two electron spin states in a
double quantum-dot system \cite{Petta2005}. As shown in Fig. \ref{fig1}a, a
double-dot system is formed by a layer of two dimensional electron gas
restrained by several electrostatic gates used to control the potentials of
individual dots and the inter-dot tunneling. Its low energy spectrum is
plotted in Fig. \ref{fig1}b. In the measurement process, the double
quantum-dot system is set in the biased regime (gray area in Fig. \ref{fig1}%
b) which can be reduced to an artificial three-level system with the
Hamiltonian%
\begin{eqnarray}
\mathcal{H}_{0} &=&E_{T}\left\vert (1,1)T_{0}\right\rangle \left\langle
(1,1)T_{0}\right\vert +E_{S}\left\vert \left( 1,1\right) S\right\rangle
\left\langle \left( 1,1\right) S\right\vert   \notag \\
&&-\varepsilon \left\vert \left( 0,2\right) S\right\rangle \left\langle
\left( 0,2\right) S\right\vert +T_{c}(\left\vert \left( 1,1\right)
S\right\rangle \left\langle \left( 0,2\right) S\right\vert   \notag \\
&&+\left\vert \left( 0,2\right) S\right\rangle \left\langle \left(
1,1\right) S\right\vert ),  \label{PhysicHamilDD}
\end{eqnarray}
where the notation $\left( n_{l},n_{r}\right) $ indicates $n_{l}$ electrons
on the "left"\ dot and $n_{r}$ electrons on the "right"\ dot, $S$ and $T$
represent spin singlet and triplet states, $\varepsilon $ and $T_{c}$ denote
the external voltage bias and tunneling amplitude between two dots. It is a
good approximation to set $E_{T}\approx E_{S}=0$, therefore Eq. (\ref%
{PhysicHamilDD}) is reduced to
\begin{eqnarray*}
\mathcal{H}_{DD} &=&-\varepsilon \left\vert \left( 0,2\right) S\right\rangle
\left\langle \left( 0,2\right) S\right\vert +T_{c}(\left\vert \left(
1,1\right) S\right\rangle \left\langle \left( 0,2\right) S\right\vert  \\
&&+\left\vert \left( 0,2\right) S\right\rangle \left\langle \left(
1,1\right) S\right\vert ).
\end{eqnarray*}

The spin measurement relies on the spin-to-charge conversion. When an
electron tunnels in or out of the right dot, the electrostatic potential in
its vicinity is changed, and the current of QPC, $I_{QPC}$ is very sensitive
to the electrostatic changes correspondingly, thus the measurement of the
changes of $I_{QPC}$ reflects the number of electrons in the right dot. In
the achieved experiment \cite{Petta2005}, the two electron spin states $%
\left\vert \left( 1,1\right) S\right\rangle $ or $\left\vert \left(
0,2\right) S\right\rangle $ can be measured with QPC since they have
different charge distribution. Here we focus on the effect of spin states
induced by QPC measurement, while the influence of the nuclear spins can be
safely neglected in case of the implementation of spin-echo or dynamical
nuclear polarization technology \cite{Petta2005,Reilly2008}.

The entire system would include the two quantum dots and the QPC near the
right dot, and the whole system Hamiltonian can be written as%
\begin{eqnarray*}
\mathcal{H} &\mathcal{=}&\mathcal{H}_{DD}+\mathcal{H}_{QPC}+\mathcal{H}%
_{int}, \\
\mathcal{H}_{QPC} &=&\sum_{U}E_{U}a_{U}^{\dagger
}a_{U}+\sum_{L}E_{L}a_{L}^{\dagger }a_{L}, \\
\mathcal{H}_{int} &=&\sum_{L,U}\{\Omega _{UL}\left\vert \left( 1,1\right)
S\right\rangle \left\langle \left( 1,1\right) S\right\vert (a_{L}^{\dagger
}a_{U}+a_{U}^{\dagger }a_{L}) \\
&&+\delta \Omega _{UL}\left\vert \left( 0,2\right) S\right\rangle
\left\langle \left( 0,2\right) S\right\vert (a_{L}^{\dagger
}a_{U}+a_{U}^{\dagger }a_{L})\}.
\end{eqnarray*}%
In the above equations, the terms can be easily understood: (1) The quantum
dots in the measurement process is a two-state system where $\left\vert
\left( 0,2\right) S\right\rangle $ and $\left\vert \left( 1,1\right)
S\right\rangle $ correspond to the case that two electrons are both in the
right dot or each electron resides in a dot. (2) QPC is described as a
standard one dimensional noninteracting electron system where $%
a_{U}^{\dagger }(a_{U})$ and $a_{L}^{\dagger }(a_{L})$ are the creation
(annihilation) operators in the upper and the lower of QPC. (3) Since the
presence of an extra electron in the right dot results in an effective
increase of the QPC barrier, the hopping amplitude in QPC can be represented
as $\Omega _{UL}$ and $\delta \Omega _{UL}$ corresponding to $\left\vert
\left( 1,1\right) S\right\rangle $ and $\left\vert \left( 0,2\right)
S\right\rangle $ states in quantum dots, respectively.

It is known that the off-diagonal density matrix elements can be destroyed
by interaction with the measurement device. We can derive the equation of
density matrix from the many-body Schr\"{o}inger equation \cite{Gurvitz1997}
or the standard technique with a Born-Markov approximation \cite{Blum}. For
convenience, the temperature of the system is assumed to be zero. The result
master equations for the entire system can be obtained:
\begin{eqnarray}
\frac{d\rho _{11}^{(n)}}{dt} &=&-D^{^{\prime }}\rho _{11}^{(n)}+D^{^{\prime
}}\rho _{11}^{(n-1)}+\frac{i}{\hbar }T_{c}(\rho _{12}^{(n)}-\rho
_{21}^{(n)}),  \label{mn00} \\
\frac{d\rho _{22}^{(n)}}{dt} &=&-D\rho _{22}^{(n)}+D\rho _{22}^{(n-1)}-\frac{%
i}{\hbar }T_{c}(\rho _{12}^{(n)}-\rho _{21}^{(n)}),  \label{mn11} \\
\frac{d\rho _{12}^{(n)}}{dt} &=&\frac{i}{\hbar }\varepsilon \rho _{12}^{(n)}+%
\frac{i}{\hbar }T_{c}(\rho _{11}^{(n)}-\rho _{22}^{(n)})-\frac{1}{2}%
(D^{^{\prime }}+D)\rho _{12}^{(n)}  \notag \\
&&+(DD^{^{\prime }})^{1/2}\rho _{12}^{(n-1)}\,.  \label{mn01}
\end{eqnarray}%
Here $\rho _{11}(t)$ and $\rho _{22}(t)$ are the probabilities of finding
the electron in the state $\left\vert \left( 0,2\right) S\right\rangle $ or $%
\left\vert \left( 1,1\right) S\right\rangle $, $\rho _{12}(t)$ and $\rho
_{21}^{\ast }(t)$\ are the off-diagonal density-matrix elements, $\rho
_{12}(t)=\rho _{21}^{\ast }(t)$. The definition $D=\mathcal{T}eV_{d}/h$ \ or
$D^{^{\prime }}=\mathcal{T}^{^{\prime }}eV_{d}/h$ is the transition rate of
electron hopping from the upper to the lower of QPC corresponding to the
quantum dot states $\left\vert \left( 1,1\right) S\right\rangle $ or $%
\left\vert \left( 0,2\right) S\right\rangle $, respectively. The QPC
transmission probability is $\mathcal{T}$ and $eV_{d}$ is the voltage bias.
The index $n$ denotes the number of electrons hopping in the QPC at time $t$.

In order to determine the influence of the QPC on the measured system, we
trace out the QPC states $\rho _{ij}=\sum_{n}\rho _{ij}^{(n)}(t)$ in Eqs. (%
\ref{mn00},\ref{mn11},\ref{mn01}), thus obtain
\begin{eqnarray}
\frac{d\rho _{11}}{dt} &=&\frac{i}{\hbar }T_{c}(\rho _{12}-\rho _{21}),
\label{M00} \\
\frac{d\rho _{22}}{dt} &=&\frac{i}{\hbar }T_{c}(\rho _{21}-\rho _{12}),
\label{M11} \\
\frac{d\rho _{12}}{dt} &=&\frac{i}{\hbar }\varepsilon \rho _{12}+\frac{i}{%
\hbar }T_{c}(\rho _{11}-\rho _{22})-\Gamma _{d}\rho _{12}.  \label{M01}
\end{eqnarray}%
The last term in the equation for the non-diagonal density matrix elements $%
\rho _{12}$ generates the exponential damping of the non-diagonal density
matrix element with the \textquotedblleft dephasing\textquotedblright\ rate
\begin{equation}
\Gamma _{d}=\frac{1}{2}(\sqrt{D}-\sqrt{D^{^{\prime }}})^{2}=(\sqrt{\mathcal{T%
}}-\sqrt{\mathcal{T}^{^{\prime }}})^{2}\frac{eV_{d}}{4\pi \hbar }.
\label{t2}
\end{equation}%
From the definition of dephasing time $T_{2}$, we know that without any
other decoherence source, $T_{2}=1/\Gamma _{d}.$ And we can find that $\rho
_{12}\rightarrow 0$ for $t\rightarrow \infty $. i.e.\ QPC measurement
destroy spin coherence of the system.

\textit{Measurement effects.} For studying the dephasing by measurement, we
first start with the case of absence of QPC. Solving the Eqs.~(\ref{M00},\ref%
{M11},\ref{M01}) for $\Gamma _{d}=0$ with the initial conditions $\rho
_{22}(0)=1$ and $\rho _{11}(0)=\rho _{12}(0)=0$ we obtain $\rho _{22}(t)=%
\frac{T_{c}^{2}\cos ^{2}(\omega t)+\varepsilon ^{2}/4}{T_{c}^{2}+\varepsilon
^{2}/4}$, where $\omega =(T_{c}^{2}+\varepsilon ^{2}/4)^{1/2}$. This is an
ordinary Bloch type time evolution of spin states in the quantum dots.

\begin{figure}[ht]
\subfigure [] {\label{a}\includegraphics[width=0.85\columnwidth]{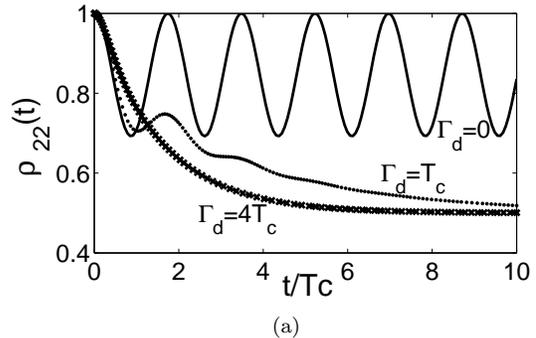}} %
\subfigure[]{\label{b}\includegraphics[width=0.85\columnwidth]{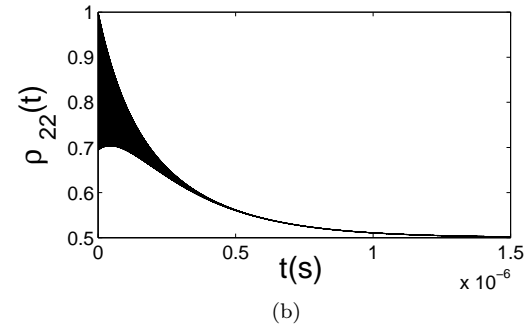}}
\caption{(a) The time-dependence of the probability to find the
double-electron state is $\left\vert \left( 0,2\right) S\right\rangle $ for $%
\protect\varepsilon =3T_{c}$. The curves correspond to different values of
the measurement induced dephasing rate: $\Gamma _{d}=0$ (solid line), $%
\Gamma _{d}=T_{c}$ (dot line), and $\Gamma _{d}=4T_{c}$ (dot-dashed line).
(b) The time-dependence of the probability to find the double-electron state
is $\left\vert \left( 1,1\right) S\right\rangle $ for real experimental
parameters.}
\label{fig2}
\end{figure}

Then the measurement device QPC is in present, the Fig. \ref{fig2}a shows
the time-dependence of the probability to find double-electron spin state in
$\left\vert \left( 1,1\right) S\right\rangle $, as obtained from the
solution of the Eqs. (\ref{M00},\ref{M11},\ref{M01}) with the initial
conditions $\rho _{22}(0)=1$ and $\rho _{11}(0)=\rho _{12}(0)=0$ for
different cases: $\varepsilon =3T_{c}$, and $\Gamma _{d}=0$ (solid line), $%
\Gamma _{d}=T_{c}$ (dot line), and $\Gamma _{d}=4T_{c}$ (dot-dashed line).
It is clear that for small $t$ the electron spins will be more localized in
its initial state $\left\vert \left( 1,1\right) S\right\rangle $ with the
increase of $\Gamma _{d}$. Since the measurement time $\Delta t$ is
inversely proportional to $\Gamma _{d}$, one could expect that the enhanced
QPC measurement will slow down the rate of transition from spin state $%
\left\vert \left( 1,1\right) S\right\rangle $ to $\left\vert \left(
0,2\right) S\right\rangle $, i.e., the frequent repeated measurement will
trap the system for small $t$. Therefore for small $t$ our results seem to
be in an agreement with quantum Zeno effect. \

Moreover, we would like to determine the spin evolution in the presence of
QPC with real experimental parameters. To proceed, we consider for
simplicity a $\delta $-potential tunnel barrier $\Omega _{UL}(x)=\frac{\hbar
^{2}b}{m^{\ast }}\delta (x)$ for the QPC, and an additional Coulomb
interaction as $\delta \Omega _{UL}=\Omega _{UL}(x)+\frac{e^{2}}{4\pi
\epsilon \epsilon _{0}a}$ due to the quantum dot transformation from spin
state $\left\vert \left( 1,1\right) S\right\rangle $ to $\left\vert \left(
0,2\right) S\right\rangle $. Using the standard scattering method the
transmission change $\Delta \mathcal{T}\,$ for different spin states can be
estimated as
\begin{equation}
\Delta \mathcal{T}=\,\mathcal{T}-\mathcal{T}^{^{\prime }}\approx \frac{e^{2}%
}{4\pi \epsilon \epsilon _{0}a}\left. \frac{\mathcal{T}(1-\mathcal{T})}{E}%
\right\vert _{E_{F}}.\;\;\;\;\;  \label{dt}
\end{equation}%
By inserting typical experimental numbers in Eq.~(\ref{dt},\ref{t2}), $%
\mathcal{T}(E=E_{F})=1/2$, $E_{F}=10\,\mathrm{meV}$, $a=200\,\,\mathrm{nm}$,
$\epsilon =13$ and $V_{d}=1\,\mathrm{mV}$, we obtain $\Delta \mathcal{T}\,/%
\mathcal{T}\approx 0.0277$ and $\Gamma _{d}\approx 1.139\times 10^{7}$ $%
\mathrm{{s}^{-1}}$. Then we can calculate the time-dependence relation of $%
\rho _{11}(t)$ as shown in Fig. \ref{fig2}b for the experimental regime: $%
\varepsilon \approx 30$ $\mathrm{\mu eV},$ $T_{c}\approx 10$ $\mathrm{\mu eV.%
}$ From Fig. \ref{fig2}b we can observe that the dephasing time is about 1 $%
\mathrm{\mu s}$ when the decoherence is only induced by the QPC measurement.

Since $\Delta \mathcal{T}\,$ is much smaller than the $\mathcal{T}$, we can
expand the Eq. (\ref{t2}) as $\Gamma _{d}=(\sqrt{\mathcal{T}}-\sqrt{\mathcal{%
T}^{^{\prime }}})^{2}\frac{eV_{d}}{4\pi \hbar }\approx \frac{eV_{d}(\Delta
\mathcal{T})^{2}}{16\pi \hbar \mathcal{T}}$. Based on the Eq. (\ref{dt}), we
can find that the dephasing time $T_{2}=1/\Gamma _{d}$ $\propto 1/\Delta
\mathcal{T}^{2}\propto a^{2},$ and $T_{2}=1/\Gamma _{d}$ $\propto 1/V_{d}.$
It is clear that the QPC position $a$ and the voltage bias $V_{d}$ determine
the dephasing time, so we can adjust $\Gamma _{d}$ or $T_{2}$ by changing
parameters $a$ and $V_{d}.$ The dephasing time from QPC measurement will
extend with the decrease of $V_{d}$ or the increase of $a$.

However, for real experimental systems when nuclear spins are not
efficiently suppressed, we can include them phenomenologically the rate $%
1/T_{2}^{env}$ which describes the intrinsic decoherence in the dot,
contributing to $1/T_{2}$. The contribution of QPC measurement to $1/T_{2}$
is calculated above as $\Gamma _{d}$, then the total decoherence rate is $%
1/T_{2}=\Gamma _{d}+1/T_{2}^{env}$. Since $\Gamma _{d}$ $\approx 1.137\times
10^{7}$ $\mathrm{{s}^{-1}}$ calculated from Eq. (\ref{t2}) and environment
induced decoherence time $T_{2}^{env}\simeq 10\,\,\mathrm{ns}$ which
recently has been measured in a double-dot setup for singlet-triplet
decoherence \cite{Petta2005}, we find that for a real experimental system in
case the nuclear spins are not efficiently suppressed we have $%
1/T_{2}^{env}\gg \Gamma _{d}$, and the effects of QPC measurement may not be
observed. Nevertheless, when the spin-echo or dynamical nuclear polarization
technology are applied \cite{Petta2005,Reilly2008} or the system is SiGe or
graphene quantum dots \cite{SiGe,Graphene}, the influence of nuclear spin is
suppressed or completely eliminated. Thus the $1/T_{2}^{env}$ will be much
small, the effect of the QPC measurement would dominate the decoherence of
the system.

\textit{Conclusion.} In summary, in this work we study in detail the effects
generated by QPC measurement of two-electron spin states in double
quantum-dot system. We give an effective Hamiltonian and derive the master
equations of the whole system. Then we calculate the time evolution of spin
states and find QPC measurement induced dephasing time $T_{2}\approx 1$ $%
\mathrm{\mu s}$. We also provide a simple and transparent description of the
enhanced QPC measurement which could trap the system for small $t$ and be
interpreted in terms of quantum Zeno effect.

\textit{Acknowledgement.} This work at USTC was funded by National
Basic Research Programme of China (Grants No. 2006CB921900 and No.
2009CB929600), the Innovation funds from Chinese Academy of
Sciences, and National Natural Science Foundation of China (Grants
No. 10604052 and No. 10874163 and No.10804104).


\end{document}